# Mechanical Model for Fiber-laden Membranes

Yogesh K. Murugesan

Alejandro D. Rey

Department of Chemical Engineering and McGill Institute of Advanced Materials

McGill University

Montreal, Quebec Canada H3A 2B2

e-mail:alejandro.rey@mcgill.ca

Date: June 23, 2009

**Abstract**

An integrated mechanical model for fiber-laden membranes is presented and representative predictions of relevance to cellulose ordering and orientation in the plant cell wall are presented. The model describes nematic liquid crystalline self-assembly of rigid fibers on an arbitrarily curved fluid membrane. The mechanics of the fluid membrane is described by the Helfrich bending-torsion model, the fiber self-assembly is described by the 2D Landau-de Gennes quadrupolar **Q**-tensor order parameter model, and the fiber-membrane interactions (inspired by an extension of the 2D Maier-Saupe model to curved surfaces) include competing curvo-philic (curvature-seeking) and curvo-phobic (curvature-avoiding) effects. Analysis of the free energy reveals three fiber orientation regimes: (a) along the major curvature, (b) along the minor curvature, (c) away from the principal curvatures, according to the competing curvo-philic and curvo-phobic interactions. The derived shape equation (normal stress balance) now includes curvature-nematic ordering contributions, with both bending and torsion renormalizations. Integration of the shape and nematic order equations gives a complete model whose solution describes the coupled membrane shape/fiber order state. Applications to cylindrical membranes, relevant to the plant cell wall, shows how growth decreases the fiber order parameter and moves the fibers' director from the axial direction towards the azimuthal orientation, eventually leading to a state of stress predicted by pure membranes. The ubiquitous $54.7^o$ cellulose fibril orientation in a cylindrical plant cell wall is shown to be predicted by the preset model when the ratio of curvo-phobic and curvo-philic interactions is in the range of the cylinder radius.

1. Introduction

The mechanical and thermodynamical description of fiber-laden membranes has implications to basic science and applications to engineering and biological systems, as in the following four phenomena. (1) The isotropic-nematic liquid crystal phase transition in 2D is a problem that continues to be investigated due to the prediction that no truly long range order can be observed in nematics [1,2]. (2) Highly packed rod-like proteins on the cell membrane systems can be described through a mesogenic self-assembly process based on the classical 2D Onsager model [3,4]. (3) The plant cell wall is a multicomponent system that includes a pectin matrix and a cellulosic fiber assembly [5]. Increase in plant cell volume during growth results in decreased curvature of the cell membrane. The cell growth process is accompanied by deposition of cellulose microfibrils in the extracellular matrix to form primary cell wall. The growth forces such as change in cell shape reorient the cellulose microfibrils within the primary cell wall which is formed during growth. As the cell wall form, the early highly curved stage orients the fibrillar cellulosic network into a mostly chiral mesophase that upon further growth eventually dissipates [6-8]. (4) Biomimetics of cell wall formation with concurrent fibrillar mesophase formation, known as the Bouligand structure, continues to provide inspiration of efficient self-assembly routes to high performance materials [6,8,9]. Necessity of an anisotropically aligned fiber layer to act as a base for formation of a defect free Bouligand structure has been numerically emphasized [10,11]. The integrated mechanical model developed for fiber-laden membranes is used to demonstrate the transitory nature of orderliness observed during development of primary cell wall. This paper seeks to contribute to the on-going theoretical effort to develop a better understanding of fiber orientation and self-assembly on soft deformable membranes, with cellulose fibrillar mesophase structure in the plant cell wall as the central long range paradigm.

Figure 1 shows a schematic that defines the main issues of fiber self-assembly on a soft deformable membrane, considered in this paper. The figure shows an ensemble of rigid fibers on a flat 2D membrane (left figure) and on a curved membrane (right figure). The membrane average (deviatoric) curvature is H (D); see section 2.1 [12-14]. The fiber orientation is defined by the director **n** and exists when the scalar order parameter S is non-zero. We consider a flat membrane (H=D=0) with a random assembly of rigid rod fibers of length L and radius R (left figure). When the fiber volume fraction θ is such that θL/2R< C, which according to Onsager 3D model [3] is close to 4, the fibers on a flat surface should be in the isotropic state. The situation we wish to describe is how under sufficient curvature, a 2D nematic state with a specific director orientation **n** may arise through curvature-mediated interactions (right figure). The main concept is that introducing mechanical bending and torsion to the membrane creates a curvature field that then may generate a 2D nematic ordering with a particular director orientation. For example, in the plant cell wall the reported cellulose fiber angle is $\tan^{-1}(2)=54.7º$ and is the fiber orientation associated with the maximum cylinder volume and the one that conforms with the stress state of $T_{\varphi\varphi} = 2T_{zz}$ where φ is the azimuthal and z is the axial coordinate in a cylindrical geometry [15]. The fluid membrane under consideration in this paper is described by the Helfrich model [16-21], and contains bending and torsion elasticity. The fibers interact with each other through excluded volume. The fiber and membrane interaction are mediated through the membrane curvature, such that both curvo-philic and curvo-phobic effects are included; in the former (latter) **n** seeks to align along high (zero) curvature directions [22]. Thus the process to be described is the coupling of nematic liquid crystal self-assembly coupled to membrane shape selection under the action of a pressure differential. Curvature-mediated interactions have been previously discussed in the literature; see for example [22-24].

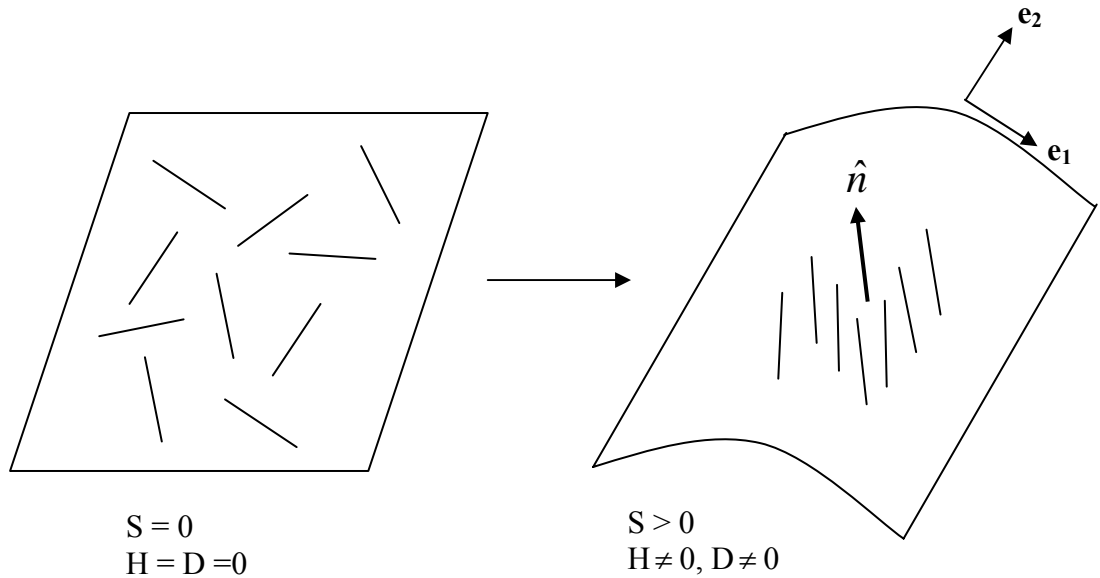

**Figure 1.** Schematic of curvature-induced 2D nematic liquid crystal self-assembly on membranes. At sufficiently low fiber packing density, the isotropic state observed on a flat membrane (left) may lead to a nematic state (right) due to curvature-mediated interactions. The nematic is described by a scalar order parameter S and a nematic director **n**. The geometry is defined by the average curvature H and the deviatoric curvature D.

The organization of this paper is as follows. Section 2 presents the membrane geometry and order parameters. A classification of surfaces and tensor bases to expand 2D tensors is introduced. The 2D quadrupolar nematic order parameters **Q** and **S** are introduced. Section 3 introduces the membrane force and torque balance equations, the

$2\times 3$ moment **M** tensor and the $2\times 3$ stress tensor **T**. Section 4 introduces the Helmholtz free energy density, which includes the Helfrich bending/torsion model, the Landau-deGennes nematic liquid crystal homogeneous energy, and the membrane-coupling energies. The free energy is similar to that presented in [22]. Section 5 analyzes curvature induced fiber ordering, and discusses three possible regimes, according to the relative magnitudes of the curvo-phobic and curvo-philic interactions: (i) alignment along maximum curvature, (ii) alignment along minimum curvature, and (iii) alignment away from the principal curvatures. Section 6 introduces the integrated membrane shape-fiber order model. Section 7 presents representative applications: spherical and cylindrical membranes. Section 8 presents the conclusions.

## 2. Geometry and Order Parameters

### 2.1 Surface Geometry

The geometry of an interface is characterized by the mean surface curvature H and the Gaussian curvature K given by [12-14]:

$$H = -\frac{1}{2}\nabla_s \cdot \mathbf{k} = \frac{1}{2}\mathbf{I}_s : \mathbf{b} = -\frac{1}{2}\mathbf{I}_s : \nabla_s \mathbf{k} = \frac{1}{2}(\upsilon_1 + \upsilon_2), K = \upsilon_1 \upsilon_2 \quad (1a,b)$$

where $\nabla_s = \mathbf{I}_s \cdot \nabla$ is the surface gradient, $\mathbf{I}_s = \mathbf{I} - \mathbf{kk}$ is the 2×2 unit surface dyadic, **k** is the unit normal, **I** is the 3×3 unit dyadic, **b** is the 2×2 symmetric surface curvature dyadic, and $\{\upsilon_m, \mathbf{e}_m\}$, m=1,2 are the eigenvalues and eigenvectors of **b**. The principal curvatures $(\upsilon_1, \upsilon_2)$ define the principal radii of curvature ($r_m$) of the surface: $\upsilon_m = -1/r_m$.

The four independent basis surface tensors are [16, 25]:

$$\{\mathbf{I}_s, \mathbf{q}, \boldsymbol{\varepsilon}_s, \mathbf{q}_1 = \mathbf{q} \cdot \boldsymbol{\varepsilon}_s\} \quad (2)$$

The matrix representations of the basis vectors in the principal frame are [16,25]:

$$\mathbf{I}_s = \begin{pmatrix} 1 & 0 \\ 0 & 1 \end{pmatrix}, \quad \mathbf{q} = \begin{pmatrix} 1 & 0 \\ 0 & -1 \end{pmatrix}, \boldsymbol{\varepsilon}_s = \begin{pmatrix} 0 & 1 \\ -1 & 0 \end{pmatrix}, \mathbf{q}_1 = \mathbf{q} \cdot \boldsymbol{\varepsilon}_s = \begin{pmatrix} 0 & 1 \\ 1 & 0 \end{pmatrix}$$

where $\boldsymbol{\varepsilon}_s$ is the surface alternator tensor. The tensor basis orthonormality yields the following results:

$$\mathbf{I}_s : \mathbf{I}_s = \mathbf{q}:\mathbf{q} = \boldsymbol{\varepsilon}_s : \boldsymbol{\varepsilon}_s = \mathbf{q}_1 : \mathbf{q}_1 = 2, \mathbf{I}_s : \mathbf{q} = \mathbf{I}_s : \boldsymbol{\varepsilon}_s = \mathbf{I}_s : \mathbf{q}_1 = \mathbf{q} : \boldsymbol{\varepsilon}_s = \mathbf{q}:\mathbf{q}_1 = \boldsymbol{\varepsilon}_s : \mathbf{q}_1 = 0,$$

$$\mathbf{I}_s \cdot \mathbf{q} = \mathbf{q}, \mathbf{I}_s \cdot \mathbf{q}_1 = \mathbf{q}_1, \mathbf{I}_s \cdot \boldsymbol{\varepsilon}_s = \boldsymbol{\varepsilon}_s, \mathbf{q} \cdot \mathbf{q}_1 = \boldsymbol{\varepsilon}_s, \mathbf{q} \cdot \boldsymbol{\varepsilon}_s = \mathbf{q}_1, \mathbf{q}_1 \cdot \boldsymbol{\varepsilon}_s = \mathbf{q} \tag{3}$$

Any 2×2 tensor $\mathbf{Z}$ can be expanded as:

$$\mathbf{Z} = \underbrace{\frac{1}{2}(\mathbf{Z}:\mathbf{I}_s)\mathbf{I}_s}_{trace} + \underbrace{\frac{1}{2}(\mathbf{Z}:\mathbf{q})\mathbf{q}}_{diagonal\ traceless} + \underbrace{\frac{1}{2}(\mathbf{Z}:\boldsymbol{\varepsilon}_s)\boldsymbol{\varepsilon}_s}_{antisymmetric} + \underbrace{\frac{1}{2}(\mathbf{Z}:\mathbf{q}_1)\mathbf{q}_1}_{symmetric\ off\text{-}diagonal} \tag{4}$$

where the subtext identifies the nature of the tensor. A symmetric 2×2 tensor diagonal in the principal coordinate frame simplifies to:

$$\mathbf{Z} = \frac{1}{2}(\mathbf{Z}:\mathbf{I}_s)\mathbf{I}_s + \frac{1}{2}(\mathbf{Z}:\mathbf{q})\mathbf{q} \tag{5}$$

Using eqn. (5) the curvature tensor b can be decomposed into a trace and a deviatoric curvature tensor: $\mathbf{b} = H\mathbf{I}_s + D\mathbf{q}$, where the deviatoric curvature D is given by: $D = (\upsilon_1 - \upsilon_2)/2$. The relation between these three curvatures is: $K = H^2 - D^2$. Depending on specific applications it is best to work with $(H,D)$ or with $(H,K)$. The magnitude of the deviatoric curvature D is a useful non-sphericity index, since for a sphere D=0. Figure 2 shows a schematic of the geometric classification of surfaces, including elliptic, cylindrical, and hyperbolic surfaces, in terms of H, D, and K [28].

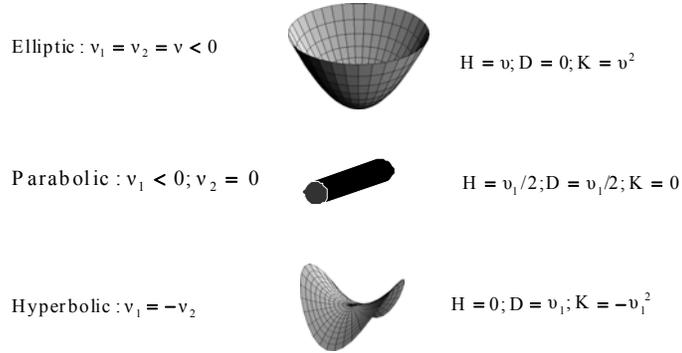

**Figure 2**. The geometric classification of surfaces. Spherical surfaces (D=0) impart no orientation preference. Cylindrical surfaces have a direction with zero curvature. Minimal (hyperbolic) surfaces have zero average curvature H=0. Adapted from [28].

## 2.2 Two Dimensional Orientation Tensor and Curvature Tensor

The characterization of nematic order is given by the $2\times 2$ symmetric traceless quadrupolar tensor order parameter **Q** [26-29]:

$$\mathbf{Q} = S\left(\mathbf{nn} - \frac{\mathbf{I}_s}{2}\right) = \mu_n \mathbf{nn} + \mu_m \mathbf{mm}, \quad \mu_n = \frac{S}{2}, \quad \mu_m = -\mu_n = -\frac{S}{2} \qquad (6)$$

where S is the scalar order parameter that measures the fiber alignment along the director **n**, $\{\mu_i\}$ are the eigenvalues of **Q**, and **n.m=0**. The range of S is $-1 \leq S \leq 1$, when S<0 the fibers are aligned along **m**, and when S=0 the fibers adopt an isotropic disordered state. In this paper we focus on non-negative values of S. Since the curvature **b** has a trace (H), the energy-coupling with nematic order involves a non-zero trace $2\times 2$ symmetric tensor **S** with non-zero eigenvalues:

$$S= Q+\frac{I_s}{2} =\left(\frac{S+1}{2}\right)nn+\left(\frac{1-S}{2}\right)mm \qquad (7)$$

The spectral decomposition of **S** and **Q** used in the free energy below, are:

$$S=\frac{1}{2}I_s +\frac{S}{2}(nn:q)q+\frac{1}{2}(nn:q_1)q_1, \quad Q=\frac{1}{2}(Q:q)q+\frac{1}{2}(Q:q_1)q_1 \qquad (8a,b)$$

### 3. Interfacial Balance Equations

In this paper we consider the isothermal mechanics of a thermodynamically closed membranes separating two bulk isotropic fluids denoted by phase 1 and phase 2, in the absence of external flow. The surface fluid model balance equations include the interfacial force balance and the interfacial torque balance [18, 30-32]. The presentation of the basic background material in Section 3 is from [30-32].

### 3.1 Interfacial Stresses and Interfacial Force Balance Equation

In this section the basic force balance is introduced and the nature of the stress tensors is defined. The $3\times 3$ stress tensors in the bulk phases are denoted by $\mathbf{T}_{b(1)}$, $\mathbf{T}_{b(2)}$, and the $2\times 3$ membrane stress tensor T is given by the sum of a tangential $\mathbf{T}_{//}$ contribution and a bending $\mathbf{T}_\perp$ contribution:

$$\mathbf{T} =\mathbf{T}_{//}+\mathbf{T}_\perp, \quad \mathbf{T}_{//}=T^{\alpha\beta}\mathbf{a}_\alpha\mathbf{a}_\beta, \quad \mathbf{T}_\perp =T^{\alpha(n)}\mathbf{a}_\alpha\mathbf{k}, \quad \alpha,\beta=1,2 \qquad (9a,b,c)$$

where // denotes the tangential plane, $\perp$ denotes the interface normal direction. The $2\times 2$ tangential stress tensor $T^{\alpha\beta}$ describes dilational (stretching) and shear stresses, while $T^{\alpha(n)}$ is the $2\times 1$ bending stress tensor. The stresses in the two bulk phases are pure pressures: $\mathbf{T}_{b(i)} =-p^{(i)}\mathbf{I}, i=1,2$.

The interfacial force balance equation is the balance between interfacial forces and the bulk stress jump [42]:

$$\nabla_s . \mathbf{T} + \mathbf{k} \cdot \left[ \mathbf{T}_{b(2)} - \mathbf{T}_{b(1)} \right] = \mathbf{0} \tag{10}$$

Projecting eqn. (10) into tangential and normal components yields the interfacial force balances [17, 18, 33, 41]:

$$\left( \nabla_s . \mathbf{T}_\perp \right) \cdot \mathbf{k} + \mathbf{T}_{//} : \mathbf{b} + \mathbf{kk} : \left[ \mathbf{T}_{b(2)} - \mathbf{T}_{b(1)} \right] = 0, \quad \left( \nabla_s \cdot \mathbf{T}_{//} \right) \cdot \mathbf{I}_s - \left( \mathbf{M} : (\nabla_s \mathbf{b}) \right) \cdot \mathbf{I}_s = 0 \tag{11a,b}$$

where eqn.(11a) is the shape equation. The term $\left( \nabla_s . \mathbf{T}_\perp \right) \cdot \mathbf{k}$ introduces the direct bending and torsion effects on the shape.

**3.2 Torque Balance Equation**

In this section we show that the interfacial torque balance equation imposes certain restrictions on the interfacial stress tensor **T**, as its components are involved in generating moments that satisfy the internal angular momentum equation. The general interfacial torque balance equation is: $\mathbf{\Gamma} = \mathbf{0}$, where $\mathbf{\Gamma}$ is the interfacial torque acting on the unit normal **k** that is generated from surface interactions; $\mathbf{\Gamma}$ has tangential and normal components:

$$\mathbf{\Gamma} = \Gamma^\alpha \mathbf{a}_\alpha + \Gamma^n \mathbf{k}, \alpha = 1, 2 \tag{12a, b}$$

No torque appears from the bulk phases because these fluids are isotropic [32]. The interfacial torque vector $\mathbf{\Gamma}$ is given by surface stress asymmetry and by gradients of surface couples stresses **N** [42]:

$$\mathbf{\Gamma} = \mathbf{T}_x + \nabla_s . \mathbf{N} \quad, \quad \mathbf{T}_x = -\mathbf{\varepsilon} : \mathbf{T} \tag{13a,b}$$

where $\mathbf{T}_x$ is the vector dual of the membrane stress tensor **T**, and **N** is the $2 \times 3$ membrane moment tensor. The dual $\mathbf{T}_x$ has tangential and normal components:

$$\mathbf{T}_x = \left( T^{\alpha(n)} \varepsilon^{\beta\lambda} a_{\lambda\alpha} \right) \mathbf{a}_\beta - \left( T^{\alpha\beta} \varepsilon_{\beta\alpha} \right) \mathbf{k} = \mathbf{\varepsilon}_s \cdot \left( \mathbf{T}_{s\perp} \cdot \mathbf{k} \right) - \left( \mathbf{\varepsilon}_s : \mathbf{T}_{s//} \right) \mathbf{k} \tag{14}$$

where $a_{\lambda\alpha}$ are the components of the surface matrix tensor, and $\varepsilon_{\beta\alpha}$ are the covariant components of the surface alternator. The normal component of the dual $\mathbf{T}_x$ is nonzero if its tangential component $\mathbf{T}_{s//}$ is asymmetric, while the tangential component of the dual $\mathbf{T}_x$ is nonzero if the bending stress $\mathbf{T}_{s\perp}$ is not zero. The second torque in eqn.(13a) is decomposed as:

$$\nabla_s \cdot \mathbf{N} = (\nabla_s \cdot \mathbf{N}) \cdot \mathbf{I}_s + (\mathbf{N} : \mathbf{b}) \mathbf{k} \tag{15}$$

Substituting eqns.(14,15) into eqn.(13a) the torque balance equation finally reduces to:

$$\boldsymbol{\varepsilon}_s \cdot (\mathbf{T}_\perp \cdot \mathbf{k}) + (\nabla_s \cdot \mathbf{N}) \cdot \mathbf{I}_s + ((\mathbf{N} : \mathbf{b}) - (\boldsymbol{\varepsilon}_s : \mathbf{T}_{//})) \mathbf{k} = 0 \tag{16}$$

Introducing the moment stress tensor $\mathbf{M} = -\mathbf{N}.\boldsymbol{\varepsilon}_s$ allows to factor out $\boldsymbol{\varepsilon}_s$ in eqn.(16) and its substitution yields the following stress tensor components:

$$\mathbf{T}_{//}^{[a]} = -\frac{1}{2}(\mathbf{M} \cdot \mathbf{b} - \mathbf{b} \cdot \mathbf{M}), \quad \mathbf{T}_\perp = -(\nabla_s \cdot \mathbf{M}) \mathbf{k} \tag{17a,b}$$

Equations (17a,b) were derived by Waxman [33].

The tangential stress tensor (eqn.(9)) is the sum of the trace $\mathbf{T}_{//}^{(tr)}$, symmetric traceless $\mathbf{T}_{//}^{(st)}$ and asymmetric $\mathbf{T}_{//}^{[a]}$ components:

$$\mathbf{T}_{//} = \mathbf{T}_{//}^{(tr)} + \mathbf{T}_{//}^{(st)} + \mathbf{T}_{//}^{(a)}, \quad \mathbf{T}_{//}^{(tr)} = \frac{1}{2}(\mathbf{T}_{//} : \mathbf{I}_s)\mathbf{I}_s, \mathbf{T}_{//}^{(st)} = \frac{1}{2}(\mathbf{T}_{//} : \mathbf{q})\mathbf{q} + \frac{1}{2}(\mathbf{T}_{//} : \mathbf{q}_1)\mathbf{q}_1, \mathbf{T}_{//}^{[a]} = \frac{1}{2}(\mathbf{T}_{//} : \boldsymbol{\varepsilon}_s)\boldsymbol{\varepsilon}_s \tag{18a,b,c}$$

Finally, the constraint $\mathbf{M}.\mathbf{b} = \mathbf{b}.\mathbf{M}$ restricts the moment stress tensor to be symmetric and of the form:

$$\mathbf{M} = \mathbf{M}^B + \mathbf{M}^T, \quad \mathbf{M}^B = \frac{1}{2}(\mathbf{M} : \mathbf{I}_s)\mathbf{I}_s, \quad \mathbf{M}^T = \frac{1}{2}(\mathbf{M} : \mathbf{q})\mathbf{q} \tag{19a,b,c}$$

where the superscript $\mathcal{B}$ denotes bending, $\mathcal{T}$ torsion. It is found that for the selected free energy (eqn.(24)), the moment tensor **M** is diagonal in the principal frame and $\mathbf{M}:\mathbf{q}_1 = 0$.

## 4. Free Energies

A number of previous works have incorporated different nematic energies into the Helfrich membrane model; the reader is referred to [22] for comprehensive discussions of previous work. In this paper we neglect spatial gradients and focus and formulate a model with curvature-driven competing interactions. Related 3D mechanical models for fibers or mesogens embedded in a 3D elastic matrix and in a 3D viscoelastic matrix have also been previously formulated [34,35,44].

The total free energy per unit area $\rho \hat{A}$ is posited to be:

$$\rho \hat{A} = \rho \hat{A}_{membrane}(\omega, H, K) + \rho \hat{A}_{fiber}(\omega, \mathbf{Q}) + \rho \hat{A}_{coupling}(\omega, \mathbf{b} \cdot \mathbf{S}) \qquad (20)$$

where $\rho$ is the density and $\omega$ is the fiber mass fraction; in this paper $\omega$ is fixed and will be omitted in what follows. The Helfrich free energy per unit area widely used to describe the elasticity of membranes reads [19]:

$$\rho \hat{A}_{membrane}(H, K) = \gamma_o + 2k_c (H - H_o)^2 + \bar{k}_c K \qquad (21)$$

where $k_c$ is bending elastic moduli, $H_o$ is the spontaneous curvature, $\bar{k}_c$ is the torsion elastic moduli; the fiber concentration on $k_c$ and $\bar{k}_c$ is neglected. The fiber contribution is given by the Landau-de Gennes expansion [22,27]:

$$\rho \hat{A}_{fiber} = \frac{a_1}{2} \mathbf{Q}:\mathbf{Q} + \frac{a_4}{4}(\mathbf{Q}:\mathbf{Q})^2 \qquad (22)$$

where $a_1 > 0, a_4 > 0$ indicating that if nematic ordering arises all it is only through the curvature effect. The coupling contribution is taken to have up to second order terms in **S**:

$$\rho \hat{A}_{coupling} = \mathbf{b} : (a_2 \mathbf{S} + a_5 \mathbf{S} \cdot \mathbf{S}) + \frac{a_3}{2}(\mathbf{b} : \mathbf{S})^2 \tag{23}$$

For $a_2 > 0$ ($a_2 < 0$) the first term $\mathbf{b} : (a_2 \mathbf{S})$ promotes fiber alignment along the principal $\mathbf{e}_1$ ($\mathbf{e}_2$) direction; the same holds for $a_5$, and hence they are denoted curvo-philic  For $a_5 > 0$, the second term $\mathbf{b} : (a_5 \mathbf{S})$ promotes nematic ordering when the fibers are oriented along the principal axes (curvo-philic). For $a_3 > 0$, the third term $a_3 (\mathbf{b} : \mathbf{S})^2 / 2$ promotes fiber orientation away from the principal axes (curvo-phobic).  The balance between these competing interactions is modified by the ambient curvature $\mathbf{b}$, since the power dependence is linear for ($a_2$, $a_5$) and quadratic for $a_3$.  The nature and meaning of the $a_2$ and $a_3$ terms has been previously discussed [22], but the role of the curvo-philic $a_5$ in creating nematic order was not.

Figure 3 shows a schematic summarizing the orientations possibilities under large and small curvatures, for the simple case of a cylinder; the scales are exaggerated to facilitate the discussion. The dot indicates the director is into the page and the horizontal rod indicates that the director is tangential to the membrane.  Since curvature-avoidance (curvo-phobic) scales with $\mathbf{b}^2$ and curvature-seeking (curvo-philic) with $\mathbf{b}$, the likely scenario of nemtaic order is axial fiber orientation at high curvature (upper left) and azimuthal orientation at low curvature (lower right).  The arrow indicates the possible existence of a change from curvo-phobic regime to a curvo-philic regime.

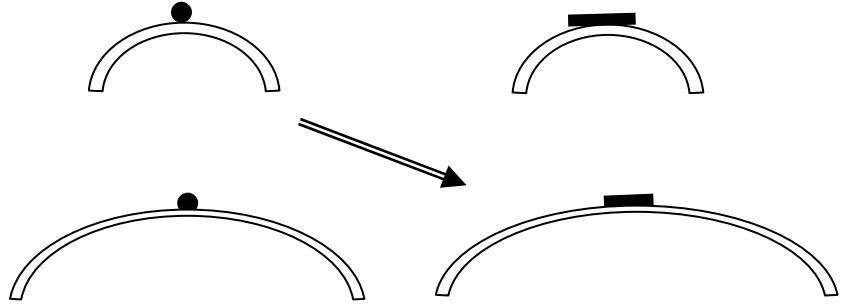

**Figure 3.** Schematic showing the fiber orientation (filled circle or line segment along the principal curvatures for a cylinder (cross sectional view and not to scale). Under competing curvature-seeking and curvature-avoiding interactions, high curvature produces axial orientation (upper left) and low curvature azimuthal orientation (lower right). The arrow indicates the expected curvature-driven orientation changes.

Using the principal curvature frame ($\mathbf{e}_1,\mathbf{e}_2$), parametrizing the director with $\mathbf{n}=(\cos\alpha, \sin\alpha)$, where $\cos\alpha=\mathbf{n}\cdot\mathbf{e}_1$, the free energy $\rho\hat{A}_{fiber} + \rho\hat{A}_{coupling}$ becomes:

$$\rho\hat{A}_{fiber}+\rho\hat{A}_{coupling}=\beta_o+\beta_2\cos^2\alpha+\beta_4\cos^4\alpha \qquad (24a,b,c,d)$$

$$\beta_o=\left(\left(a_2+\frac{a_5}{2}\right)+\frac{a_3}{2}H\right)H-\left((a_2+a_5)D+a_3HD\right)S+\left(\frac{a_1}{4}+\frac{a_5}{2}H+\frac{a_3}{2}D^2\right)S^2+\frac{a_4}{16}S^4$$

$$\beta_2=2(a_2+a_3H-a_3DS+a_5)DS, \quad \beta_4=2a_3D^2S^2$$

This energy vanishes for flat isotropic states: D=H=S=0. The quartic contribution ($\beta_4\cos^4\alpha$) is necessary to observe fiber alignment away from the principal axes. For a spherical surface (D=0) the free energy is independent of $\alpha$. The form of the free energy is identical to that describing the interfacial tension of nematic-substrate interfaces [36-39].

The formulation of the fiber and coupling energies (eqns.(22,23)) is suggested by the 2D Maier-Saupe liquid crystal model for curved interfaces. For flat interfaces the 2D Maier-Saupe potential that describes liquid crystal self-assembly is [40]:

$$\Phi_Q(\mathbf{u}) = -U\mathbf{Q}:\mathbf{uu} \tag{25}$$

where U is the nematic potential and **u** is the fiber orientation vector. For curved interfaces its generalization is:

$$\Phi(\mathbf{u}) = \Phi_Q(\mathbf{u}) + \Phi_b(\mathbf{u}) = -U\mathbf{Q}:\mathbf{uu} - V\ell\mathbf{b}:\mathbf{uu} \tag{26}$$

where $\ell$ is characteristic coupling length between curvature and orientation and V is curvature potential. Note that

$$\mathbf{Q}:\mathbf{uu} = \mathbf{Q}:\left(\mathbf{uu} - \frac{\mathbf{I}_s}{2}\right), \quad \mathbf{b}:\mathbf{uu} = \mathbf{b}:\left(\mathbf{uu} - \frac{\mathbf{I}_s}{2}\right) + H \tag{27a,b}$$

The term in eqn.(26) gives rise to terms of the form **Q:Q** while the second to terms that include **b:S**. The **Q**-potential is derived from an internal energy E(**Q**) as follow:

$$\Phi_Q = \frac{\partial E(\mathbf{Q})}{\partial \mathbf{Q}}:\mathbf{uu} = -U(\mathbf{Q}):(\mathbf{uu}), \quad E(\mathbf{Q}) = -\frac{1}{2}U(\mathbf{Q}:\mathbf{Q}) \tag{28a,b}$$

According to the Maier-Saupe model, the entropy is:

$$TS_{ent} = -U\mathbf{Q}:\mathbf{Q} + kT\ln Z = 2E + kT\ln Z \tag{29}$$

where the single molecule orientation partition function Z is:

$$Z = \int e^{-(\Phi_Q + \Phi_b)/k_B T} d\mathbf{u} \tag{30}$$

Using eqns.(29) and (30), the molar Helmholtz free energy $A_m$ of the system is found to be:

$$A_m = -N_A E - N_A k_B T\ln Z = \frac{N_A}{2}U\mathbf{Q}:\mathbf{Q} - N_A k_B T\ln Z \tag{31}$$

If one expands the lnZ term in eqn.(31) in a Taylor series, terms of the form appearing in the phenomenological eqn.(22,23) will be included; the exact calculation is beyond the scope of the paper but the sought-after molecular motivation for eqn.(24) is demonstrated.

## 5. Curvature-induced Fiber Ordering

In this section we identify the three possible regimes of fiber orientation with respect to the main curvature directions ($e_1$, $e_2$) that arise when the curvature tensor **b** deviates from the zero tensor. Given the different curvature effects on fiber orientation identified in conjunction of eqn.(23) and their different weights with increasing H and D, several fiber orientation transitions will arise.

For a given geometry, the preferred fiber orientation ($\alpha$) and fiber order (S) is found by minimizing the Helmholtz free energy density $\rho \hat{A}$ :

$$\rho \frac{\partial \hat{A}}{\partial S} = \frac{a_4}{4} S^3 + \left( \frac{a_1}{2} + a_5 H \right) S + \left( a_2 + a_3 H + a_5 + a_3 DS \cos 2\alpha \right) D \cos 2\alpha = 0 \qquad (32a,b)$$

$$\rho \frac{\partial \hat{A}}{\partial \alpha} \rightarrow \left( \frac{\beta_2}{2\beta_4} + \cos^2 \alpha \right) \sin \alpha \cos \alpha = 0$$

where the second equation is re-written in a more revealing format. In terms of the fiber director $\alpha$ there are three equilibria states: (a) orientation along major curvature, $\alpha=0$; (b) along the minor curvature, $\alpha=\pi/2$; and (c) oblique, $\cos^2 \alpha = -\beta_2 / 2\beta_4$, as follows.

(a) Orientation along the major curvature. Here $\mathbf{n}=\mathbf{e}_1$ and the scalar order parameter S satisfies a cubic:

$$\alpha = 0 \quad , \quad \frac{a_4}{4} S^3 + \left( \frac{a_1}{2} + a_5 H + a_3 D^2 \right) S + \left( a_2 D + a_3 HD + a_5 D \right) = 0 \qquad (33)$$

The necessary conditions are

$$\frac{\beta_2}{2\beta_4} = \frac{\left( a_2 + a_5 + a_3 (H - D S) \right)}{2 a_3 D S} > 0, \qquad \beta_2 = 2(a_2 + a_3 H - a_3 DS + a_5) DS < 0 \qquad (34)$$

Since $\beta_4 < 0$, $\beta_2 < 0$ the minimum energy corresponds to orientation along the largest principal curvature ($\alpha=0$). For a cylinder this is the azimuthal direction.

(b) Orientation along the minor curvature. Here $\mathbf{n}=\mathbf{e}_2$ and the scalar order parameter S satisfies a cubic:

$$\alpha = \pi/2 \quad, \quad \frac{a_4}{4}S^3 + \left(\frac{a_1}{2}+a_5 H + a_3 D^2\right)S - \left(a_2 D + a_3 HD + a_5 D\right) = 0 \tag{35}$$

The necessary conditions are

$$\frac{\beta_2}{2\beta_4} = \frac{\left(a_2 + a_5 + a_3(H - D S)\right)}{2a_3 D S} > 0, \qquad \beta_2 = 2\left(a_2 + a_3 H - a_3 DS + a_5\right)DS > 0 \tag{36}$$

Since $\beta_4 > 0$, $\beta_2 > 0$ the minimum energy corresponds to orientation along the smallest principal curvature ($\alpha=\pi/2$). For a cylinder this is the axial direction.

(c) Oblique orientation along the minor curvature. Here $\mathbf{n}=\mathbf{e}_2$ and the scalar order parameter S satisfies a cubic:

$$\cos^2 \alpha = -\frac{\beta_2}{2\beta_4} = -\frac{\left(a_2 + a_5 + a_3 H - a_3 D S\right)}{2a_3 D S} \quad, \quad \left(\frac{a_1}{2}+a_5 H\right)S + \frac{a_4}{4}S^3 = 0 \tag{37a,b}$$

The necessary conditions $\left(-2\beta_4 < \beta_2 < 0, \beta_4 > 0\right)$

$$-2a_3 D^2 S^2 < \left(a_2 + a_3 H - a_3 DS + a_5\right)DS < 0, \quad \beta_4 = 2a_3 D^2 S^2 > 0 \tag{38}$$

For this particular model (eqns. (24)) we find a null-potential $\Omega$ for the oblique case:

$$\rho\frac{\partial \hat{A}}{\partial S} = \underbrace{\frac{\partial \beta_0}{\partial S} + \frac{\partial \beta_2}{2\partial S} + \frac{\partial \beta_4}{4\partial S}}_{eqn.(37b)} + \underbrace{\left(\frac{\partial \beta_2}{\partial S} + \frac{\partial \beta_4}{\partial S} + \frac{\partial \beta_4}{\partial S}\frac{\cos 2\alpha}{2}\right)\frac{\cos 2\alpha}{2}}_{eqn.(37a):\Omega=0} \tag{39}$$

since we have the restriction:

$$\frac{\frac{\partial \beta_2}{\partial S}+\frac{\partial \beta_4}{2\partial S}}{\frac{\partial \beta_4}{\partial S}}=\frac{\beta_2}{2\beta_4}$$

This indicates that ordering under oblique conditions is the result of a competition of a competition between the Landau-deGennes terms and the curvo-philic $a_5$ term.

Next we discuss transitions and assume all the energy coefficients $\{a_i\}>0$. For $\beta_4>0$, orientation transitions from oblique to principal curvature directions occurs for $D\neq 0$ as follows. (a) Oblique-major curvature ($e_1$) transition: $as\ -\beta_2 \to 2\beta_4, \alpha \to 0$. In this case $\beta_2$ becomes sufficiently negative to balance $\beta_4$. At the transition when $\alpha \to 0$ the curvatures obey:

$$D\sqrt{-\frac{(2a_1+4a_5 H)}{a_4}}+H=-\frac{(a_2+a_5)}{a_3}<0 \tag{40}$$

This transition will occur at sufficiently small curvatures so that the effect of $a_2$ and $a_5$ prevail (see discussion below eqn.(23)).

(b) Oblique-minor curvature ($e_2$): $as\ \beta_2 \to 0, \alpha \to \pi/2$. This transition to orientation along the minimum curvature occurs as the tension introducing $\cos^2\alpha$ in the free energy vanishes (see eqn.(24c)) and this can happen only at sufficiently large H and D. At the transition the curvatures obey:

$$H-D\sqrt{-\frac{(2a_1+4a_5 H)}{a_4}}=-\frac{(a_2+a_5)}{a_3} \tag{41}$$

Next we demonstrate an application of these results (eqns.(40,41)) for the case of a cylinder of radius R, when D=H=-1/2R, relevant to cellulose fibrils in plant cell walls.

Using eqns.(40,41), we find that the critical transition radii $R_1$ and $R_2$ for oblique/major curvature and oblique/minor curvature, respectively, obey:

$$R_1 = \frac{a_3}{2(a_2+a_5)}\left\{1+\sqrt{-\frac{2\left(a_1-\frac{a_5}{R_1}\right)}{a_4}}\right\} = \frac{a_3}{2(a_2+a_5)}(1+S) \tag{42}$$

$$R_2 = \frac{a_3}{2(a_2+a_5)}\left\{1-\sqrt{-\frac{2\left(a_1-\frac{a_5}{R_2}\right)}{a_4}}\right\} = \frac{a_3}{2(a_2+a_5)}(1-S) =< R_1 \tag{43}$$

where we used eqn.(37b) for S. Figure 4 summarizes these findings. A small cylinder radius induces axial orientation while larger radius induces azimuthal orientation, and the interval over which the intermediate oblique state exists is:

$$R_1 - R_2 = \frac{a_3}{2(a_2+a_5)}\left(\sqrt{\frac{\left(\frac{a_5}{R_1}-a_1\right)}{a_4/2}} + \sqrt{\frac{\left(\frac{a_5}{R_2}-a_1\right)}{a_4/2}}\right) \tag{44}$$

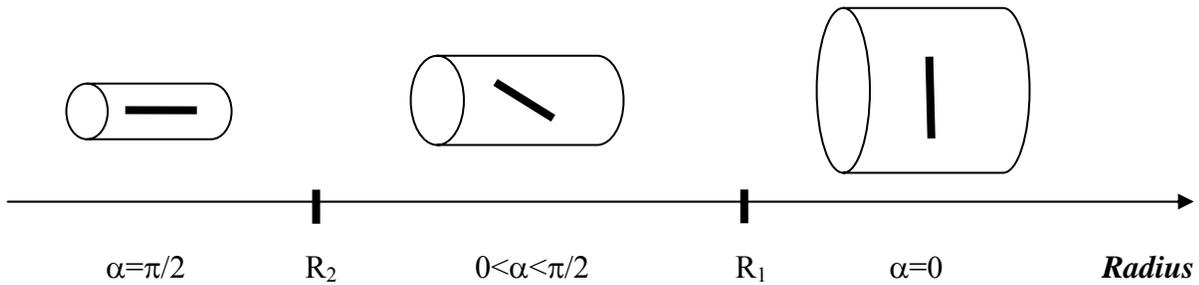

$\alpha=\pi/2$     $R_2$     $0<\alpha<\pi/2$     $R_1$     $\alpha=0$     *Radius*

**Figure 4**. Schematic of curvature-induced fiber orientation (thick line segment) on a cylinder of increasing radius R. At small (large) radius, the fibers align along the axial (azimuthal) direction. At intermediate curvatures, the oblique state minimizes the free energy (eqns.(23,37)).

These regimes and transitions are rationalized by identifying the dimensionless numbers for the model. Form the free energy we find the bare internal length scale $\ell_o$ associated with the fiber orientation:

$$\ell_o = \frac{a_3}{a_2 + a_5} = \frac{curvo-phobic\ energy}{curvo-philic\ energy} \tag{45}$$

As shown by eqns. (42,43), due to the presence of the scalar order parameter S the bare length is renormalized

$$\ell = \ell_o(1 \pm S) \tag{46}$$

Hence the dimensionless number P that controls the fiber orientation is the ratio of the internal length scaled to the external length scale:

$$P = \frac{internal\ length}{external\ length} = \frac{\ell}{R} = \frac{\ell_o(1 \pm S)}{R} \tag{47}$$

where R is the characteristic radius of curvature or cylinder radius. We find the following regimes and transitions.

(i) Curvo-phobic Regime: The internal length scale is much larger than the cylinder radius

$$P \to \infty, \ell \gg R \tag{48}$$

The curvo-phobic effect dominates and the state is selected by minimizing the $a_3$ term:

$$\rho \hat{A}_{coupling} = \underbrace{\mathbf{b}:(a_2 \mathbf{S} + a_5 \mathbf{S} \cdot \mathbf{S})}_{negligible} + \underbrace{\frac{a_3}{2}(\mathbf{b}:\mathbf{S})^2}_{minimized} \tag{49}$$

(ii) Axial-Oblique transition: the transition occurs at

$$P = \frac{\ell_o(1-S)}{R} = 1 \tag{50}$$

(iii) Intermediate Oblique Regime: here we have competing curvo-phobic and curvo-philic interactions and P is in the range:

$$\frac{\ell_o(1-S)}{R} < P < \frac{\ell_o(1+S)}{R} \tag{51}$$

All the energies in eqn.(23) are important.

(iv) Oblique-Azimuthal transition: the transition occurs at

$$P = \frac{\ell_o(1+S)}{R} = 1 \tag{52}$$

(v) Curvo-philic Regime: The internal length scale is much smaller than the cylinder radius

$$P \to 0, \ell \ll R \tag{53}$$

The curvo-philic effect dominates and the state is selected by minimizing the $a_3$ term:

$$\rho \hat{A}_{coupling} = \underbrace{\mathbf{b}:(a_2\mathbf{S} + a_5\mathbf{S}\cdot\mathbf{S})}_{minimized} + \underbrace{\frac{a_3}{2}(\mathbf{b}:\mathbf{S})^2}_{negligible} \tag{54}$$

In partial summary, curvature re-orients the fiber director along a preferred direction that minimizes ordering and coupling energies, according to the signs of the tension coefficients $\beta_2$ and $\beta_4$ (eqn.(24)). For this particular model, the oblique orientation nullifies part of the potential (denoted $\Omega$ in eqn.(60)). Oblique orientation is favored by sufficient large average curvature H and non-zero deviatoric D curvature. When D=0, fiber orientation is degenerate. For minimal surfaces (H=0) nematic ordering exists only when fibers are parallel to the principal axes. Table I summarizes the possible ordering-orientation regimes of the fibers on the three different classes of surfaces (elliptic, parabolic, and hyperbolic), found using eqns. (32). For a representative elliptic surface we use the sphere (D=0). Each surface class may exhibit orientation along the minor curvature (MI), major curvature (MA) or along a oblique direction (OB), with the exception of minimal surfaces (H=0).

TABLE I: Fiber orientation and order on elliptic, parabolic and Hyperbolic surfaces

|  | Elliptic surface | | | Parabolic surface | | | Hyperbolic surface | | |
|---|---|---|---|---|---|---|---|---|---|
|  | MI | MA | OB | MI | MA | OB | MI | MA | OB |
| S | $S = \sqrt{\dfrac{-4\left(\dfrac{a_1}{2}+a_5 H\right)}{a_4}}$ | | | $\dfrac{a_4}{4}S^3$ $+\left(\dfrac{a_1}{2}+a_1 H+a_3 H^2\right)S$ $+(a_2 H+a_3 H^2+a_4 H)=0$ | $\dfrac{a_4}{4}S^3$ $+\left(\dfrac{a_1}{2}+a_1 H+a_3 H^2\right)S$ $-(a_2 H+a_3 H^2+a_4 H)=0$ | $S = \sqrt{\dfrac{-4\left(\dfrac{a_1}{2}+a_5 H\right)}{a_4}}$ | $\dfrac{a_4}{4}S^3$ $+\left(\dfrac{a_1}{2}+a_3 D^2\right)S$ $+(a_2 D+a_3 D)=0$ | $\dfrac{a_4}{4}S^3$ $+\left(\dfrac{a_1}{2}+a_3 D^2\right)S$ $-(a_2 D+a_3 D)=0$ | 0 |
| $\cos^2\alpha$ | $0 \le \cos^2\alpha \le 1$ (degenerate) | | | 1 | 0 | $-\dfrac{(a_2+a_4+a_3 H-a_5 HS)}{2a_3 HS}$ | 1 | 0 | NA |

## 6. Integrated Membrane Shape-Fiber Order Model

The membrane shape-fiber ordering determination requires the integration of the shape equation and the nematic structure equation. Derivation of the model without fibers has been presented in [30-32]. Here we follow the same sequence and procedure and to avoid length repetitions we quote the main results and direct the reader to [30-32] for additional details.

To find the shape equation we follow previous work [30, 32] and formulate: (i) the membrane tension γ, (ii) the elastic moment tensor **M**, and (iii) the stress tensor **T** and finally (iv) the static shape equation.

(i) According to the surface Euler equation [13], the Helmholtz free energy per unit mass $\hat{A}$ is given by:

$$\hat{A} = \omega\left(\mu_f - \mu_m\right) + \frac{\gamma}{\rho} \tag{55}$$

where $\left(\mu_f, \mu_m\right)$ are the fiber and membrane chemical potentials per unit mass, ρ is the membrane mass density, and γ is the membrane tension. Using eqns.(20,55) we find

$$\gamma = \gamma_o + 2k_c\left(H - H_o\right)^2 + \bar{k}_c K + \beta_o + \beta_2 \cos^2\alpha + \beta_4 \cos^4\alpha \tag{56}$$

where $\gamma_o$ is the tension at zero curvature (H=K=0) under zero order (S=0).

(ii) The moment tensor **M** is found from the variation of energy with curvature (26):

$$\mathbf{M} = \frac{\rho \partial \hat{A}}{\partial \mathbf{b}} = \mathbf{M}^B + \mathbf{M}^T, \quad \mathbf{M}^B = \frac{B}{2}\mathbf{I}_s, \quad \mathbf{M}^T = \frac{T}{2}\mathbf{q} \tag{57}$$

where the static bending and torsion are

$$B = C_1 + 2C_2 H + \frac{1}{2}\left[\Omega + a_5 S^2\right], \quad T = \left\{-2C_2 D + \Omega(\mathbf{Q} : \mathbf{q})\right\} \tag{58}$$

and where the coefficients $\{C_1, C_2, \Omega\}$ are:

$$C_1 = \left(\frac{\rho \partial \hat{A}_{membrane}}{\partial H}\right)_K = 4k_c(H - H_o), \quad C_2 = \left(\frac{\rho \partial \hat{A}_{membrane}}{\partial K}\right)_H = \overline{k}_c \tag{59a,b}$$

$$\Omega = \frac{1}{\mathbf{Q} : \mathbf{q}_1}\left(\frac{\rho \partial \hat{A}}{\partial \mathbf{b}} : \mathbf{q}_1\right) = a_2 + a_5 + a_3 H + a_3 SD \cos 2\alpha \tag{60}$$

The new bending/torsion contributions due to nematic ordering are:

$$B_{nematic} = \frac{1}{2}\left(\cos^2 \alpha + \frac{\beta_2}{2\beta_4} + a_5 S^2\right), \quad T_{nematic} = \left(\cos^2 \alpha + \frac{\beta_2}{2\beta_4}\right) S \cos 2\alpha \tag{61}$$

Since S and $\alpha$ are function of **b**, the bending/torsion is nonlinear in H and D. The nematic bending and torsion coefficients $B_{nematic}, T_{nematic}$ are largest under fiber orientation along the maximum curvature, while the torsion is positive for $\alpha=0$, negative for $\alpha=\pi/2$ and zero for oblique states. Using eqns.(62a,b) we find the following stress contributions (see eqn.(64) below):

$$\mathbf{M} \cdot \mathbf{b} = \left(\frac{BH}{2} + \frac{TD}{2}\right)\mathbf{I}_s + \left(\frac{TH}{2} + \frac{BD}{2}\right)\mathbf{q}, \quad \frac{1}{2}\mathbf{M} : \mathbf{b} = \left(\frac{BH}{2} + \frac{TD}{2}\right)\mathbf{I}_s \tag{62, 63}$$

Equation (63) shows that **M·b** is diagonal in the principal curvature frame and that the stress tensor is symmetric (see eqn.(17a)).

(iii) To find the elastic tangential stress **T** we use a simple variation of the Helmholtz free energy $\mathcal{A} = \int \rho \hat{A}(\rho, \mathbf{b}) dS$ with respect to areal mass density $\rho$ and curvature tensor **b**. The result of this variational calculation is the following elastic stress tensor **T** [30]:

$$\mathbf{T} = \gamma \mathbf{I}_s - \mathbf{M} \cdot \mathbf{b} - (\nabla_s \cdot \mathbf{M}) \mathbf{k} \tag{64}$$

in agreement with previous variational calculations (16,20). Using eqns.(62,63,64) the components of the elastic interface stress tensor **T** are found to be:

$$\mathbf{T}_{//}^{(tr)} = \left( \gamma - \left( \frac{\mathcal{B}H}{2} + \frac{\mathcal{T}D}{2} \right) \right) \mathbf{I}_s, \quad \mathbf{T}_{//}^{(st)} = -\left( \frac{\mathcal{T}H}{2} + \frac{\mathcal{B}D}{2} \right) \mathbf{q}, \quad \mathbf{T}_\perp = -\{ \mathbf{I}_s \cdot [\nabla_s \cdot (\mathbf{M})] \} \mathbf{k} \tag{65a,b,c}$$

The tangential stress tensor is symmetric and diagonal in the principal curvature frame.

(iv) The static shape equation, obtained by introducing eqns.(65a,b,c) into (16), gives the static Helfrich shape equation [16,19, 43]:

$$\Delta P = 2H\left( \gamma - \frac{1}{2}(\mathcal{B}H + \mathcal{T}D) \right) - D((\mathcal{B}D + \mathcal{T}H)) - \nabla_s \cdot \{ \mathbf{I}_s \cdot [\nabla_s \cdot (\mathbf{M})] \} \tag{66}$$

where $\Delta P = -\mathbf{k}\mathbf{k} : [\mathbf{T}_{b(2)} - \mathbf{T}_{b(1)}]$ is the pressure difference between the two adjacent bulk phases, here considered to be isotropic fluids; see (18,19) for extensive discussions. The terms on the right hand side of eqn.(66) correspond to pressures due to dilation, shear, bending, and torsion.

Integrating the shape and fiber order equation then gives the generic coupled shape-structure equation:

$$\begin{cases} \Delta P - 2H\left( \gamma(S,\alpha) - \frac{1}{2}(\mathcal{B}(S,\alpha)H + \mathcal{T}(S,\alpha)D) \right) + D((\mathcal{B}(S,\alpha)D + \mathcal{T}(S,\alpha)H)) \\ + \nabla_s \cdot \{ \mathbf{I}_s \cdot [\nabla_s \cdot (\mathbf{M}(S,\alpha))] \} = 0 \\ \frac{\rho \partial \hat{A}(H,D,S,\alpha)}{\partial S} = 0, \quad \frac{\rho \partial \hat{A}(H,D,S,\alpha)}{\partial \alpha} = 0 \end{cases} \tag{67a,b,c}$$

In the general case numerical solutions are required to solve the coupled membrane shape-fiber orientation equations (67).

**7. Application: Fiber-laden Spherical and Cylindrical Membranes**

As mentioned in the introduction cellulose fibril organization in the plant cell wall belong to a class of mechanical problems that involves fiber packing and orientation on curved substrates [44]. In classical plant science the plant cell wall is described by a cylinder under pressure while the cellulose fibrils wound around the cylinder usually following a geodesic with an oblique angle close to 54.7º [44]. The fiber arrangement is designed to provide reinforcement for the large hoop stresses that appear in pressurized cylinders. Hence we seek to use the previously presented model to establish what particular solution of our generalized model describes cellulose organization. In addition we also consider the spherical geometry since this geometry (D=0) nullifies curvo-phobic and curvo-philic competing interactions. For the cylinder we restrict the analysis to fiber orientation away from the principal curvatures, when eqn.(39) holds . We use $H_o=0$ in eqn.(56). For both applications we adopt the following the calculation sequence:

$$\underbrace{\text{select membrane geometry}\left(sign\ of\ K\right)}_{Fig.(1)} \to \underbrace{\text{calculate fiber order }(\mathbf{Q})}_{eqns.(32)} \to$$
$$\underbrace{\text{determine membrane shape}(\mathbf{b})}_{eqn.(66)} \to \underbrace{\text{evaluate surface tangential stress tensor }(\mathbf{T}_{//})}_{eqns.(65)} \to$$
$$\underbrace{\text{compare with Helfrich model}}_{reference[16]}$$

(a) Spherical Membrane

As an example of elliptic shape (K>0) we use the sphere. For a spherical shape D=0, $\beta_2 = \beta_4 = 0$ (see eqns.(24a,b,c,d)) and the fiber orientation eqn.(32a) is satisfied for any α. The fiber **Q**-parameter is

$$Q = \sqrt{\frac{\left(\frac{2a_5}{a_1 R} - 1\right)}{a_4 / 2a_1}} \left(\mathbf{nn} - \frac{\mathbf{I}_s}{2}\right), \mathbf{n} = (\cos\alpha, \sin\alpha), \ 0 < \alpha < \pi/2 \tag{68, 69}$$

As R increases, the fiber order parameter S decreases for any angle $\alpha$. The shape eqn.(66) is cubic in R:

$$\Delta P R^3 + (2\gamma_o) R^2 - \left(\frac{3a_2}{2} + \frac{a_5}{2} + \frac{a_5 a_1}{a_4}\right) R + \left(\frac{a_3}{2} + \frac{2a_5^2}{a_4}\right) = 0 \tag{70}$$

and gives R as a function of $\Delta P$. At the maximum radius $R_{max} = 2a_5/a_1$, when S=0, we recover the Laplace equation, where the minimum pressure drop magnitude is:

$$-\Delta P_{min} = \frac{\gamma_o a_1}{a_5} \tag{71}$$

Using eqns.(37,58-60,65) the tangential stress tensor components are:

$$\mathbf{T}_{//} = \begin{pmatrix} T_{\theta\theta} & T_{\theta\varphi} \\ T_{\varphi\theta} & T_{\varphi\varphi} \end{pmatrix} = \begin{pmatrix} \gamma - \frac{H\mathcal{B}}{2} & 0 \\ 0 & \gamma - \frac{H\mathcal{B}}{2} \end{pmatrix} \tag{72}$$

$$T_{\theta\theta} = T_{\varphi\varphi} = \gamma_0 - \left(\frac{3a_2}{2} + \frac{a_1 a_5}{a_4} + \frac{a_5}{2}\right)\frac{1}{2R} + \left(\frac{a_3}{2} + \frac{2a_5^2}{a_4}\right)\frac{1}{2R^2} = \frac{-\Delta P R}{2} \tag{73}$$

At the maximum radius $R_{max} = 2a_5/a_1$ we recover the classical result [44]: $\gamma_0 = -\Delta P R/2$. The net fiber effect is to renormalize the bare tension $\gamma_o$, and the addition of new 1/R terms. Comparing the response of fiber-laden ($R_f$) and a clean ($R_o$) sphere to a pressure drop $\Delta P$ we find from the $T_{\varphi\varphi}$ stress the tube hardening equation:

$$\gamma_0 \left(\frac{1}{R_f} - \frac{1}{R_o}\right) = \left(\frac{3a_2}{4} + \frac{a_1 a_5}{2a_4} + \frac{a_5}{4}\right)\left(\frac{1}{R_f^2}\right) - \left(\frac{a_3}{4} + \frac{2a_5^2}{2a_4}\right)\left(\frac{1}{R_f^3}\right) \tag{74}$$

The main fiber contribution to the sphere hardening arises through the bending and tension renormalization due to:

$$B_{nematic} = \frac{1}{2}\left(a_2 + a_3\left(-\frac{1}{R}\right) + a_5(1+S^2)\right) \tag{75}$$

$$\rho\hat{A}_{fiber} + \rho\hat{A}_{coupling} = \left(\left(a_2 + \frac{a_5}{2}\right) + \frac{a_3}{2}\left(-\frac{1}{R}\right)\right)\left(-\frac{1}{R}\right) + \left(\frac{a_1}{4} + \frac{a_5}{2}\left(-\frac{1}{R}\right)\right)S^2 + \left(\frac{a_4}{16}\right)S^4 \tag{76}$$

and involves the Landau- deGennes energy ($a_1$,$a_4$), and coupling energies ($a_2$,$a_3$,$a_5$).

(b) Cylindrical Membrane

The parabolic shape (K=0) is the cylinder. For the cylindrical shape H=D. The fiber **Q**-parameter in terms of S and cosα is specified by:

$$S = \sqrt{\frac{2\left(\frac{a_5}{Ra_1} - 1\right)}{a_4/a_1}}, \quad \cos^2\alpha = \frac{\left(\frac{Ra_2}{a_3}\right)\left(1 + \frac{a_5}{a_2}\right) + \frac{1}{2}(S-1)}{S} \tag{77}$$

Nematic order S is a decreasing function of increasing R and the fiber orientation α decreases with increasing R. Due to the physical interval 0<S<1, the cylinder radius R is constrained to:

$$R_{min}(S=1) = \frac{2a_5}{a_4 + 2a_1} < R < R_{max}(S=0) = \frac{a_5}{a_1} \tag{78}$$

Here we are interested only in the cases that lead to 0<S<1. The values of S at the limits of fiber angle α are:

$$S(\alpha = \pi/2) = 1 - \frac{2(a_2 + a_5)}{a_3}R, \quad S(\alpha = 0) = \frac{2(a_2 + a_5)}{a_3}R - 1 \tag{79}$$

On the other hand, the radii range for oblique orientation is $R_2<R<R_1$ (see eqn.(44)). The regime of interest is when $R_{min}<R_2<R_{max}$ and $R_1<R_{max}$, that is, as the radius increases from $R_2$ to $R_1$, the fiber angle decreases from axial to azimuthal and the scalar order parameter decreases from $S(\alpha = \pi/2)$ to $S(\alpha = 0)$. The increase of R is driven by changes in $\Delta P$, and hence we must use the shape eqn.(66).

Using eqns.(65) in a cylindrical geometry (H=D) it is found that the tangential membrane stress tensor is diagonal in the principal frame:

$$\mathbf{T}_{//} = \begin{pmatrix} T_{\varphi\varphi} & T_{\varphi z} \\ T_{z\varphi} & T_{zz} \end{pmatrix} = \begin{pmatrix} \gamma\text{-H}(\mathcal{B}+\mathcal{T}) & 0 \\ 0 & \gamma \end{pmatrix} \tag{80}$$

where

$$T_{\varphi\varphi} = \gamma_o - \left(\frac{k_c}{2}\right)\frac{1}{R^2} + \left\{-\left(\frac{a_1^2}{4a_4} + \frac{(a_2+a_5)^2}{2a_3}\right) + \left(\frac{a_5}{4}\right)\frac{1}{R} + \left(\frac{a_5^2}{4a_4}\right)\frac{1}{R^2}\right\} = -\Delta PR \tag{81}$$

$$T_{zz} = \gamma_o + \left(\frac{k_c}{2}\right)\frac{1}{R^2} + \left\{-\left(\frac{a_1^2}{4a_4} + \frac{(a_2+a_5)^2}{2a_3}\right) + \left(\frac{a_5}{4} + \frac{a_1 a_5}{2a_4}\right)\frac{1}{R} + \left(-\frac{a_5^2}{4a_4}\right)\frac{1}{R^2}\right\} \tag{82}$$

where the brackets include the fiber contributions. At $R_{max}(S=0) = a_5/a_1$ we recover the classical result:

$$T_{\varphi\varphi} = \gamma_o - \left(\frac{k_c}{2}\right)\frac{1}{R^2} = -\Delta PR \quad , \quad T_{zz} = \gamma_o + \left(\frac{k_c}{2}\right)\frac{1}{R^2} \tag{83a,b}$$

Note than in this purely 2D model, the shape equation does not involve $T_{zz}$ but from 3D force balance [15] one finds $T_{\varphi\varphi} = 2 T_{zz}$, such that when using eqn.(83b) one gets $T_{zz} = -\Delta PR/2$. Furthermore when $T_{\varphi\varphi} = 2 T_{zz}$ the classical results holds in our model:

$$-\Delta PR^2 = \mathcal{B} \tag{84}$$

Using eqn.(66), for a given $\Delta P$, the selected R is found to satisfy a cubic equation:

$$\Delta PR^3 + \left\{\gamma_o - \left(\frac{a_1^2}{4a_4} + \frac{(a_2+a_5)^2}{2a_3}\right)\right\}R^2 + \left(\frac{a_5}{4}\right)R + \left\{-\left(\frac{k_c}{2}\right) + \left(\frac{a_5^2}{4a_4}\right)\right\} = 0 \tag{85}$$

and the key prediction is that the cylinder growth re-orients the fibers away from the axial direction and decreases the nematic order. The net fiber effect is to renormalize the bare tension $\gamma_o$, the bending coefficient $k_c$ and the addition of new 1/R terms. Comparing the response of fiber-laden ($R_f$) and a clean ($R_o$) cylinder to an inflation pressure $\Delta P$ we find from the azimuthal $T_{\varphi\varphi}$ stress the tube hardening equation:

$$\gamma_o\left(\frac{1}{R_f}-\frac{1}{R_o}\right)-\left(\frac{k_c}{2}\right)\left(\frac{1}{R_f^3}-\frac{1}{R_o^3}\right)=\left(\frac{a_1^2}{4a_4}+\frac{(a_2+a_5)^2}{2a_3}\right)-\left(\frac{a_5}{4}\right)\frac{1}{R_f^2}-\left(\frac{a_5^2}{4a_4}\right)\frac{1}{R_f^3} \tag{86}$$

The main fiber contribution to the tube hardening arises through the bending and tension renormalization due to:

$$B_{fiber}=\frac{a_5}{2}S^2, \gamma_{fiber}=\left(\left(a_2+\frac{a_5}{2}\right)\left(-\frac{1}{2R}\right)+\frac{a_3}{2}\left(-\frac{1}{2R}\right)^2\right) \tag{87}$$

and involve order energy ($a_1,a_4$), and coupling energies ($a_2,a_3,a_5$).

Using eqns.(77) we can estimate the model conditions that lead to the biological fibrillar orientation given by $\tan^2\alpha=2$ [44] and obtain the relation between the dimensionless number $P_o$ and the order parameter S:

$$P_o=\frac{\ell_o}{R}=\frac{6}{3-S} \tag{88}$$

Hence the parameteric P range that may describe the cell wall condition is:

$$P_{min}=\frac{\ell_o}{R_{max}}=2<P<P_{max}=\frac{\ell_o}{R_{min}}=3 \tag{89}$$

where we used equs.(77,78). For $\tan^2\alpha=2$, the stress condition $T_{\varphi\varphi}=2T_{zz}$ now yields:

$$\underbrace{(\Delta P)R^3-2k_c}_{membrane}+\underbrace{\left(\frac{18a_5}{\ell_o^2}\right)R^3-18\frac{a_5R^2}{\ell_o}+\frac{9a_5}{2}R}_{membrane/fiber}=0 \tag{90}$$

and indicates the renormalization of the membrane condition by the competing curvo-phobic and curvo-philic effects.

## 8. Conclusions

A model that describes curvature-induced fiber orientation in fluid membranes was formulated using a Helmholtz energy that integrates the Helfrich bending/torsion energy, the nematic Landau-deGennes fiber orientation energy, and the fiber-membrane couplings (eqn.(24)). The model contains competing curvo-philic and curvo-phobic interactions defined by the magnitude of an internal length scale $\ell_o$ (eqn.(45)). Under large curvatures, the curvo-phobic effect dominates and the fibers tend to orient along zero curvature directions, at large curvature the curvo-philic effect dominate and the fibers orient along the maximum principal curvature and at intermediate curvature the orientation is oblique in the principal curvature frame (see Figure 4). The Helmholtz free energy of the fiber-laden membrane was used to derive the stress and moment tensors and the shape equation (section 6). Integrating the orientation and ordering potentials with the shape equation gives a model that describes the coupled fiber orientation, fiber order, and membrane shape for a given pressure differential (eqns.(67)). Applications of the integrated model to spherical and cylindrical shapes provide a window on the different roles of average and deviatoric curvatures (eqn.(68)). For spheres, the fiber orientation is degenerate and increasing the radius just randomizes the fibers (eqn.(69)). For cylinders, the highly ordered axially aligned fibers that are found at small radii (curvo-phobic effect), leads to gradual re-orientation towards the azimuthal direction (curvo-philic effect) and lower fiber order (eqns.(77)). The azimuthal stress balances the pressure drop and for sufficiently large cylinders it converges to the Helfrich prediction (eqns.(83)).

The parametric range, given in terms of the cylinder radius R and internal length scale $\ell_o$, that describes cellulose orientation in the plant cell wall at 55º is found to be between 2 and 3 (eqn.(88)).

**Acknowledegements**

Financial support for this work was provided by the Natural Sciences and Engineering Research Council (NSERC) of Canada (A.D.R). Y.K.M. is the recipient of post-graduate scholarships from McGill Engineering Doctoral Awards (MEDA) and the Eugenie Ulmer Lamothe fund of the Department of Chemical Engineering of McGill University.